\documentclass[11pt]{article}

\usepackage{graphicx}
\usepackage{epsfig}
\usepackage{amsmath}
\usepackage{amsfonts}
\usepackage{amssymb}
\usepackage{epstopdf}
\usepackage{amsthm}
\usepackage[usenames]{color}
\usepackage{array}
\usepackage{xcolor}
\usepackage{float}
\usepackage{arydshln}
\usepackage{subfigure}

\usepackage[
      colorlinks=true,
      linkcolor=blue,
      urlcolor=blue,
      filecolor=blue,
      citecolor=red,
      pdfstartview=FitV,
      pdftitle={},
      pdfauthor={},
      pdfsubject={},
      pdfkeywords={},
      pdfpagemode=None,
      bookmarksopen=true
]{hyperref}

\usepackage{hyperref}

\def\cA{{\cal A}}
\def\cR{{\cal R}}
\def\cS{{\cal S}}
\def\cT{{\cal T}}
\def\fg{{\mathfrak g}}

\def\be{\begin{equation}}
\def\ee{\end{equation}}
\def\beq{\begin{eqnarray}}
\def\eeq{\end{eqnarray}}

\begin{document}

\addtolength{\headsep}{1cm}

\begin{centering}
  \textbf{\Large{Gravitational collapse with rotating thin shells and cosmic censorship}}

\vspace{1cm}
  \large{Jorge V. Rocha}

\vspace{0.5cm}
\begin{minipage}{.9\textwidth}
  \small \it
    \begin{center}
Centro Multidisciplinar de Astrof\'isica -- CENTRA,\\
Instituto Superior T\'ecnico, University of Lisbon,\\
Avenida Rovisco Pais 1, 1049-001 Lisboa, Portugal\\
\bigskip 
\tt jorge.v.rocha@tecnico.ulisboa.pt
    \end{center}
\end{minipage}

\end{centering}

\vspace{1cm}

\begin{abstract}
Gravitational collapse of matter in the presence of rotation is a mostly unexplored topic but it might have important implications for cosmic censorship. Recently a convenient setup was identified to address this problem, by considering thin matter shells at the interface between two equal angular momenta Myers-Perry spacetimes in five dimensions.
This note provides more details about the matching of such cohomogeneity-1 spacetimes and extends the results obtained therein to arbitrary higher odd dimensions. It is also pointed out that oscillatory orbits for shells in asymptotically flat spacetimes can be naturally obtained if the matter has a negative pressure component.
\end{abstract}



\bigskip
\section{Introduction}

The study of gravitational collapse in General Relativity has led to important insights into the fate of massive stars, the formation of black holes, and the nature of spacetime singularities~\cite{Gundlach:2007gc,Joshi:2012mk}. In this respect, most investigations over the yeas have benefitted from the assumption of spherical symmetry. Spherical symmetry and rotation are closely tied, so clearly this assumption is not a good approximation for a highly spinning system.

From the point of view of the (weak) cosmic censorship conjecture~\cite{Penrose:1969pc}, the restriction to non-rotating collapses is a strong limitation. This hypothesis is at the basis of General Relativity as a complete, self-consistent theory. In essence, it claims that any physical singularities generated by the evolution of generic and regular initial data with reasonable matter content must be hidden inside event horizons. Thus, if the conjecture is true, generic gravitational collapses leading to curvature singularities must produce black holes instead of naked singularities. What is understood as `reasonable matter' and `generic' is not explicitly defined, which leaves some room to explore. Nevertheless, it is quite obvious that spherically symmetric collapses do not configure `generic' scenarios. In fact, there are known examples of non-rotating gravitational collapses that produce naked singularities~\cite{Joshi:2012mk}. It is therefore highly desirable to go beyond spherical symmetry.

There are by now several studies of non-spherically symmetric collapse in numerical relativity, starting with work done in the 1980's~\cite{Nakamura:1981,Stark:1985da}. However, the available literature is very limited in what concerns analytic solutions describing gravitational collapse of matter in the presence of rotation. Until recently the only investigations in this vein were performed in $2+1$ dimensions, where the low dimensionality of the problem provides a crucial simplification~\cite{Crisostomo:2003xz, Mann:2008rx, Vaz:2008uv} (or in systems with assumed symmetries that reduce the problem down to $2+1$ dimensions). Ref.~\cite{Lindblom:1974bq} considered the gravitational collapse of a rotating shell in four spacetime dimensions but relied on a slow rotation approximation.

A recent development in this topic~\cite{Delsate:2014iia} identified a context in which gravitational collapse with rotating matter can be tackled analytically, without the restriction to two spatial dimensions. The construction relies on the consideration of a thin shell (for which the Darmois-Israel formalism~\cite{Israel:1966rt, Darmois} for matching two spacetimes is adopted) lying at the interface between two equal angular momenta Myers-Perry black hole solutions in odd dimensions $D\geq5$. The restriction to odd dimensions and to this family of spacetimes is similar in spirit to the consideration of spherical symmetry: despite carrying angular momentum, this class of solutions depends on a single radial coordinate~\cite{Bizon:2005cp,Kunduri:2006qa}.

The framework developed in Ref.~\cite{Delsate:2014iia}, which enables an exact study of the effect of rotation on the gravitational collapse, was restricted to the lowest dimension to which it applies, namely $D=5$. Here I review the formalism, providing more details about the matching procedure, and extend the results to arbitrary higher (odd) dimensions. This is the content of sections~\ref{sec:matching}, \ref{sec:energyconditions} and~\ref{sec:EOM}. The case of a rotating matter shell falling from rest at infinity is analyzed in section~\ref{sec:collapse}. It is proven that this process respects cosmic censorship as long as the rest mass of the shell is positive. Section~\ref{sec:oscillatory} is dedicated to the investigation of shells with a negative pressure component and it is pointed out that this can naturally lead to oscillatory motion. Finally, I conclude in section~\ref{sec:conc} with some remarks.

\section{Matching Cohomogeneity-1 Spacetimes in Higher Odd Dimensions\label{sec:matching}}

Consider the celebrated Myers-Perry family of exact vacuum solutions of the Einstein equations in arbitrary $D\geq4$ dimensions~\cite{Myers:1986un}. These geometries describe stationary rotating black holes and are characterized by a mass parameter and $N+1$ spin parameters, where $N\equiv\lfloor(D-3)/2\rfloor$.\footnote{Here, $\lfloor x \rfloor$ denotes the integer part of $x$. The number of independent spin parameters is equal to the number of available orthogonal rotation planes in $D$ dimensions, which is $N+1$ with this definition.}

When $D$ is odd (so that $D=2N+3$) and all the independent spin parameters are set to the same value, the isometry group gets enhanced from $\mathbb{R}\times U(1)^{N+1}$ to $\mathbb{R}\times U(N+1)$. In this case  --- which we will assume from now onwards --- coordinates $y^\mu$ can be found which reflect this large amount of symmetry, with the line element depending on a single (radial) coordinate~\cite{Kunduri:2006qa}. (The explicit form of the metric will be given below, in Eq.~\eqref{eq:metric}.) Spacetimes with this property are commonly referred to as cohomogeneity-1 spacetimes.

In this manuscript only asymptotically flat spacetimes are considered. A non vanishing cosmological constant can be easily included~\cite{Delsate:2014iia}. We adopt natural units in which $G=c=1$.

\subsection{Background geometries}

As motivated above, the interior and exterior spacetimes (identified with subscripts $-$ and $+$, respectively) will be taken to be equal angular momenta Myers-Perry solutions in $D=2N+3$ dimensions, with $N$ an integer. Thus, the metrics describing the two geometries acquire the same form but are characterized, in general, by different parameters $(M_\pm,a_\pm)$. In coordinates $y^\mu=(t,r,\psi,x^a)$ the metrics can be written as follows~\cite{Kunduri:2006qa}:
\beq
  ds_\pm^2 = g_{\mu\nu} dy^\mu dy^\nu  &=&  - f_\pm(r)^2 dt^2 + g_\pm(r)^2 dr^2 + r^2 \widehat{g}_{ab} dx^a dx^b \nonumber\\
 && + h_\pm(r)^2 \left[ d\psi + A_a dx^a - \Omega_\pm(r) dt \right]^2  \,,
\label{eq:metric}
\eeq
where
\begin{flalign}
&  g_\pm(r)^2 = \left( 1 - \frac{2M_\pm}{r^{2N}} + \frac{2M_\pm a_\pm^2}{r^{2N+2}} \right)^{-1}\,, 
\label{eq:metricfuncs1}\\
&  h_\pm(r)^2=r^2\left(1+\frac{2M_\pm a_\pm^2}{r^{2N+2}} \right)\,, 
    \qquad \Omega_\pm(r)=\frac{2M_\pm a_\pm}{r^{2N} h_\pm(r)^2}\,, 
    \qquad  f_\pm(r)=\frac{r}{g_\pm(r) h_\pm(r)}\,.
\label{eq:metricfuncs2}
\end{flalign}
%
The $x^a$ are $2N$ coordinates on the complex projective space $CP^N$, $\widehat{g}_{ab}$ is the Fubini-Study metric on such a manifold and $A=A_a dx^a$ denotes the associated K\"ahler potential. Expressions for $\widehat{g}_{ab}$ and $A_a$ can be obtained iteratively in $N$ (see Ref.~\cite{Hoxha:2000jf}) but they are not explicitly required for this construction. The coordinates $y^\mu$ cover the whole spacetime and run over $\{t,r,\psi,x^a\}$. The angular coordinate $\psi$ parameterizes an $S^1$ fiber and has periodicity $2\pi$ (see Ref.~\cite{Kunduri:2006qa} for more details).

The event horizon occurs at the largest real root of $g^{-2}$ (if it exists), and its spatial sections have the geometry of a homogeneously squashed $(2N+1)$-sphere. The mass ${\cal M}_\pm$ and angular momentum ${\cal J}_\pm$ of the spacetime are given by~\cite{Kunduri:2006qa}
\be
  {\cal M}_\pm = \frac{\cA_{2N+1}}{4\pi} \left(N+\frac{1}{2} \right) M_\pm\,, \qquad\quad
  {\cal J}_\pm = \frac{\cA_{2N+1}}{4\pi}(N+1)M_\pm a_\pm \,.
  \label{eq:charges}
\ee
where $\cA_{2N+1}$ is the area of a unit $S^{2N+1}$.

\subsection{Junction conditions}

Now I will match two spacetimes with line elements of the form~\eqref{eq:metric} across a timelike hypersurface $\Sigma$ defined parametrically by the equations $t=\cT(\tau)$ and $r=\cR(\tau)$, where $\tau$ is for the proper time for an observer comoving with the hypersurface. It will prove convenient to work in a corotating frame~\cite{Poisson:2004}, which is achieved by changing coordinates according to
\be
  d\psi \longrightarrow d\psi + \Omega_\pm(\cR) dt\,.
\ee
The metrics then become
\beq
  ds_\pm^2  &=&  - f_\pm(r)^2 dt^2 + g_\pm(r)^2 dr^2  + r^2 \widehat{g}_{ab} dx^a dx^b \nonumber\\
  && + h_\pm(r)^2 \left[ d\psi + A_a dx^a - (\Omega_\pm(\cR)-\Omega_\pm(r)) dt \right]^2\,.
\label{eq:metric_corot}
\eeq

The standard Darmois-Israel junction conditions~\cite{Israel:1966rt,Darmois},
\begin{flalign}
  &[[\fg_{ij}]]=0  \qquad \Rightarrow \qquad \fg_{ij}^{(+)} = \fg_{ij}^{(-)} \equiv \fg_{ij}\,,
\label{eq:junction1}\\
  &[[k_{ij}]] - \fg_{ij} [[k]] = - 8\pi G \cS_{ij} \,,
\label{eq:junction2}
\end{flalign}
follow from the requirement that the metric of the total spacetime is continuous across the thin shell and determines the form of the surface stress-energy tensor $\cS_{ij}$ that sources possible curvature singularities along $\Sigma$. Here, $\fg_{ij}$ represents the induced metric on the hypersurface $\Sigma$, while $k_{ij}$ denotes the extrinsic curvature and $k=\fg^{ij}k_{ij}$ is its trace. For any given tensorial quantity $C_{ij\dots}$, its jump across the hypersurface $\Sigma$ is defined by
\be
  [[C_{ij\dots}]]\equiv C_{ij\dots}^{(+)}-C_{ij\dots}^{(-)}\,.
\ee
%

\subsubsection{Induced metric and first junction condition}

The induced metric on the hypersurface $\Sigma$ is given by
\be
\fg_{ij}^{(\pm)} dy^i dy^j= -d\tau^2+h_{\pm}(\cR)^2[d\psi+A_a dx^a]^2+\cR^2 \widehat{g}_{ab} dx^a dx^b\,.
\ee
The coordinates $y^i$ run over $\{\tau,\psi,x^a\}$. The form above, fixing $\fg_{\tau\tau}=-1$, imposes
\be
f(\cR)^2 \dot{\cT}^2 - g(\cR)^2 \dot{\cR}^2=1\,,
\ee
where the overdot stands for $d/d\tau$. This condition determines $\dot{\cT}$ in terms of $\dot{\cR}$, up to an overall sign\footnote{Outside the event horizon (and inside the Cauchy horizon) one must have $\dot{\cT}>0$. However, between the event and Cauchy horizons $\dot{\cT}$ can have either sign.}. The contravariant components of the induced metric are given by
\be
\fg^{ij}=
\left( \begin{array}{c:c:c}
-1 & 0 & 0 \\ \hdashline
0 & \frac{1}{h(\cR)^2}+\frac{\widehat{g}^{ab}A_aA_b}{\cR^2} & -\frac{\widehat{g}^{ab}A_a}{\cR^2} \\ \hdashline
0 & -\frac{\widehat{g}^{ab}A_b}{\cR^2} & \frac{\widehat{g}^{ab}}{\cR^2} \end{array} \right),
\label{eq:invgij}
\ee
where $\widehat{g}^{ab}$ is the inverse of $\widehat{g}_{ab}$, i.e., $\widehat{g}^{ab}\widehat{g}_{bc}=\delta^a_c$.

The first junction condition~\eqref{eq:junction1} immediately gives
\be
h_+(\cR)=h_-(\cR)\equiv h(\cR)   \qquad  \Rightarrow  \qquad   M_+a_+^2 = M_-a_-^2\,.
\label{eq:Ma2}
\ee
This implies that a such a rotating cohomogeneity-1 exterior (with $M_+a_+\neq0$) cannot be continuously joined with a flat interior ($M_-=0$). In other words, in the presence of rotation our matched solutions require the interior geometry to be that of a black hole (or a naked singularity).

\subsubsection{Extrinsic curvature}

A basis of vectors tangent to $\Sigma$ is provided by $\{e_\tau,e_\psi,e_a\}$, where
\be
e_\tau^\mu=(\dot{\cT},\dot{\cR},0,\overbrace{0,\dots,0}^{\substack{2N}})\,, \quad
e_\psi^\mu=(0,0,1,\overbrace{0,\dots,0}^{\substack{2N}})\,,
\ee
and $e_a$ (with $a=1,\dots,2N$) are vectors generating the $CP^N$ component of the geometry.

The unit normal to $\Sigma$ is then given by
\be
n_\mu = f g (-\dot{\cR},\dot{\cT},0,\overbrace{0,\dots,0}^{\substack{2N}})\,.
\ee
The metric, normal vector and the $2N+2$ tangent vectors $e_i$ determine the exterior curvature through $k_{ij}=e^\mu_i e^\nu_j \nabla_\mu n_\nu$. one obtains straightforwardly
\beq
k_{\tau\tau}^{(\pm)} &=& -\frac{h}{\cR}\beta_\pm'\,, \qquad\quad\,
k_{\tau\psi}^{(\pm)} = -\frac{h^3 \Omega_\pm'}{2\cR}\,, \qquad\quad
k_{\tau a}^{(\pm)} = k_{\tau\psi}^{(\pm)}A_a\,, \\
k_{\psi\psi}^{(\pm)} &=& \frac{h^2 h'\beta_\pm}{\cR}\,, \qquad\quad
k_{\psi a}^{(\pm)} = k_{\psi\psi}^{(\pm)}A_a\,, \qquad\quad\;
k_{ab}^{(\pm)} = k_{\psi\psi}^{(\pm)}A_aA_b + \widehat{g}_{ab}h\beta_\pm\,,
\eeq
where a prime stands for $d/d\cR$ and
\be
\beta_\pm \equiv \epsilon_\pm f_\pm \sqrt{1+g_\pm^2\dot{\cR}^2}\,.
\label{eq:beta}
\ee
I have inserted above the factor $\epsilon_\pm \equiv \text{sign}(\dot{\cT})$, directly related with the sign of the radial component of the unit normal, $n^r$. The choice of these signs constrains the allowable trajectories in the maximal analytic extension of the metrics considered~\cite{Boulware:1973,Gao:2008jy}. However, they will have no influence on the shell's equation of motion that is derived in Section~\ref{sec:EOM}.

Using Eq.~\eqref{eq:invgij} one obtains the trace of the extrinsic curvature,
\be
k^{(\pm)} =  \frac{1}{\cR^{2N+1}}[\cR^{2N}h\beta_\pm]' = \frac{1}{\cR^{D-2}}[\cR^{D-3}h\beta_\pm]'\,.
\ee
Note that none of the extrinsic curvature components depend explicitly on the number of the spacetime dimensions, although there is an implicit dependence introduced through the definition of the metric functions $g, f, h, \Omega$. However, the trace does depend explicitly on $D$.

\subsection{Second junction condition and stress-energy tensor}

The second junction condition~\eqref{eq:junction2} dictates the form of the stress-energy tensor to be that of an imperfect fluid,
\be
\cS_{ij} = (\rho+P)u_iu_j + P\fg_{ij} + 2\varphi\, u_{(i}\xi_{j)} + \Delta P\, \cR^2 \widehat{g}_{ij}\,,
\label{eq:Sij}
\ee
where
\be
\widehat{g}_{ij} \equiv \widehat{g}_{ab} \frac{dx^a}{dy^i} \frac{dx^b}{dy^j}\,.
\ee
Here, $u=u^i \partial_i=\partial_\tau$ is the normalised fluid velocity (assumed to be corotating with the shell), and I define for convenience the normalised vector $\xi=\xi^i \partial_i=h^{-1}\partial_\psi$. The covariant components of these vectors are given by
\be
u_i = (-1,0,\overbrace{0,\dots,0}^{\substack{2N}})\,, \qquad\qquad \xi_i=h(0,1,\overbrace{A_a}^{\substack{2N}})\,.
\ee
The quantity $\varphi$ represents the intrinsic momentum of the fluid, while $\Delta P$ denotes the pressure anisotropy.

Specifically, the various components of the stress-energy tensor, obtained from the second junction condition, are the following:
\begin{flalign}
& \rho= - \frac{[[\beta(\cR)]]}{8\pi G\,\cR^{2N+1}} \frac{d}{d\cR}\big[\cR^{2N}h(\cR)\big]\,, \label{eq:rho}\\
& P= \frac{h(\cR)}{8\pi G\,\cR^{2N+1}} \frac{d}{d\cR}\big[\cR^{2N}[[\beta(\cR)]]\big]\,, \label{eq:P}\\
& \varphi= - \frac{h(\cR)^2}{16\pi G\,\cR} [[\Omega'(\cR)]]\,, \label{eq:varphi}\\
& \Delta P= \frac{[[\beta(\cR)]]}{8\pi G} \frac{d}{d\cR}\left[\frac{h(\cR)}{\cR}\right]\,. \label{eq:DeltaP}
\end{flalign}
It can be shown that such a stress-energy is covariantly conserved. Explicitly, the conservation equations, $\nabla^i\cS_{ij}=0$, reduce to just a pair of independent equations,
\begin{flalign}
&(P+\rho)h'\cR\dot{\cR} + h\left[ 2N(P+\Delta P+\rho)\dot{\cR}+\cR\dot{\rho} \right]=0\,, \label{eq:conserv1}\\
&2\varphi\, h'\cR\dot{\cR} + h\left( 2N\varphi\dot{\cR}+\cR\dot{\varphi} \right)=0\,. \label{eq:conserv2}
\end{flalign}
The latter equation is equivalent to
\be
\frac{d}{d\tau}\left[\varphi\,\cR^{2N}h(\cR)^2\right]=0\,,
\ee
which is automatically satisfied when $\varphi$ is given by~\eqref{eq:varphi}\footnote{To see this, one must use the definitions of the metric functions $\Omega$ and $h$.}. Similarly, Eq.~\eqref{eq:conserv1} is obeyed if the energy density, pressure and pressure anisotropy are given by Eqs.~(\ref{eq:rho}), (\ref{eq:P}) and (\ref{eq:DeltaP}), respectively.

\section{Energy Conditions\label{sec:energyconditions}}

The standard energy conditions are most conveniently expressed in terms of the eigenvalues of the stress-energy tensor~\eqref{eq:Sij}. These are obtained as the coefficients $\lambda_n$, with $n=0,\dots,2N+1$, such that
\be
\det[\cS_{ij}-\lambda_n \fg_{ij}]=0\,.
\ee
The eigenvalues can be determined without explicit knowledge of the Fubini-Study metric $\widehat{g}_{ab}$ since the determinant evaluates to
\begin{flalign}
&\det[\cR^2(P+\Delta P -\lambda_n) \widehat{g}_{ab}]
\det \left( \begin{array}{cc}
\rho+\lambda_n & \varphi h(\cR) \nonumber\\ 
\varphi h(\cR) & (P-\lambda_n) h(\cR)^2
\end{array} \right) \\
& = -\det[\widehat{g}_{ab}] \left(\cR^2(P+\Delta P -\lambda_n)\right)^{2N} \left[\lambda_n^2+(\rho-P)\lambda_n+(\varphi^2-\rho P)\right] \,.
\end{flalign}
Therefore the eigenvalues are given by
\beq
\lambda_0 &=& \frac{P-\rho}{2} - \sqrt{\left(\frac{P+\rho}{2}\right)^2-\varphi^2}\,, \\
\lambda_1 &=& \frac{P-\rho}{2} + \sqrt{\left(\frac{P+\rho}{2}\right)^2-\varphi^2}\,, \\
\lambda_\alpha &=& P+\Delta P\,, \qquad \alpha=2,\dots,2N+1. 
\eeq
Note the eigenvector associated to $\lambda_0$ is timelike as long as $\rho+P\geq0$. Moreover, the $2N$ eigenvalues $\lambda_\alpha$ are degenerate and the expressions for the eigenvalues are independent of the dimensionality $D=2N+3$.

The weak energy conditions can now be formulated simply as~\cite{Wald:1984rg}
\be
WEC_0 \equiv -\lambda_0 \geq 0\,, \qquad
WEC_1 \equiv \lambda_1-\lambda_0 \geq 0\,, \qquad
WEC_\alpha \equiv \lambda_\alpha-\lambda_0 \geq 0\,.
\label{eq:WEC}
\ee
In the non-rotating case the intrinsic momentum $\varphi$ and the pressure anisotropy $\Delta P$ both vanish and the weak energy conditions reduce to the familiar relations: $\rho\geq0$ and $\rho+P\geq0$.
The less restrictive null energy condition simply omits the first inequality.

\section{Radial Equation of Motion for the Shell\label{sec:EOM}}

Having determined the stress-energy tensor from the junction conditions, we can still impose an equation of state $P=P(\rho)$. In what follows I shall choose a linear equation of state for simplicity but more general equations of state can be considered, e.g., polytropic.

Adopting an equation of state of the form $P=w\rho$, and inserting expressions~(\ref{eq:rho}) and (\ref{eq:P}) into this relation one can easily integrate the equation to obtain
\be
[[\beta(\cR)]]= - \frac{m_0^{1+\frac{2N+1}{2N}w}}{\cR^{2N(1+w)}h(\cR)^w}\,,
\ee
where $m_0$ is an integration constant with dimensions of mass\footnote{Indeed, the rest mass of the shell is given by
$ \frac{\cA_{2N+1}}{4\pi} \left(N+\frac{1}{2} \right) m_0$, at least for the $w=0$ case I will focus on below.
The numerical coefficients that appear in this expression are exactly the same as in Eq.~\eqref{eq:charges}.}.

Using the definition~\eqref{eq:beta}, the above solution can be rearranged and cast in the standard form of an equation of motion for a classical particle moving in a one-dimensional radial potential,
\be
  \dot \cR^2 + V_{\rm eff}(\cR) =0\,,
\label{eq:motion}
\ee
where the effective potential $V_{\rm eff}$ is given by
\beq
  V_{\rm eff}(\cR) &=& \frac{1}{2}\left(g_+^{-2}+g_-^{-2}\right) - \frac{h^2}{4\cR^2} \left(\frac{m_0^{1+\frac{2N+1}{2N}w}}{\cR^{2N(1+w)}h^w}\right)^2  - \frac{\cR^2}{4h^2} \frac{\left(g_+^{-2}-g_-^{-2}\right)^2}{\left(\frac{m_0^{1+\frac{2N+1}{2N}w}}{\cR^{2N(1+w)}h^w}\right)^2} \nonumber\\
  &=& 1 + \frac{2Ma^2}{\cR^{2N+2}} - \frac{M_++M_-}{\cR^{2N}}  -\frac{1}{4} \left(\frac{m_0}{\cR^{2N}}\right)^{2+\frac{2N+1}{N}w} \left(1+\frac{2Ma^2}{\cR^{2N+2}}\right)^{1-w} \nonumber\\
  && - \left(\frac{M_+-M_-}{m_0}\right)^2 \left(\frac{\cR^{2N}}{m_0}\right)^{\frac{2N+1}{N}w} \left(1+\frac{2Ma^2}{\cR^{2N+2}}\right)^{w-1}.
\label{eq:potential}
\eeq
This generalizes the effective potential obtained by Delsate et al.~\cite{Delsate:2014iia} for five dimensions (corresponding to $N=1$) to arbitrary higher odd dimensions\footnote{The inclusion of a cosmological constant term in the Einstein equations --- and consequently in the solution Eqs.~(\ref{eq:metric}--\ref{eq:metricfuncs2}) --- simply adds the terms $\frac{\cR^2}{\ell^2}+\frac{2Ma^2}{\ell^2 r^{2N}}$ to the effective potential.}. Clearly, classically allowed regions for shell trajectories are those for which $V_{\rm eff}$ is non-positive and the turning points (if they exist) occur whenever $V_{\rm eff}=0$.

\section{Collapse From Infinity Onto a Pre-existing Black Hole\label{sec:collapse}}

Having in mind applications of this machinery to tests of the cosmic censorship conjecture, I now turn the attention to the class of thin shell solutions describing collapses from infinity. 

\begin{figure}[t]
\centering
\subfigure{\includegraphics[width=0.49\textwidth]{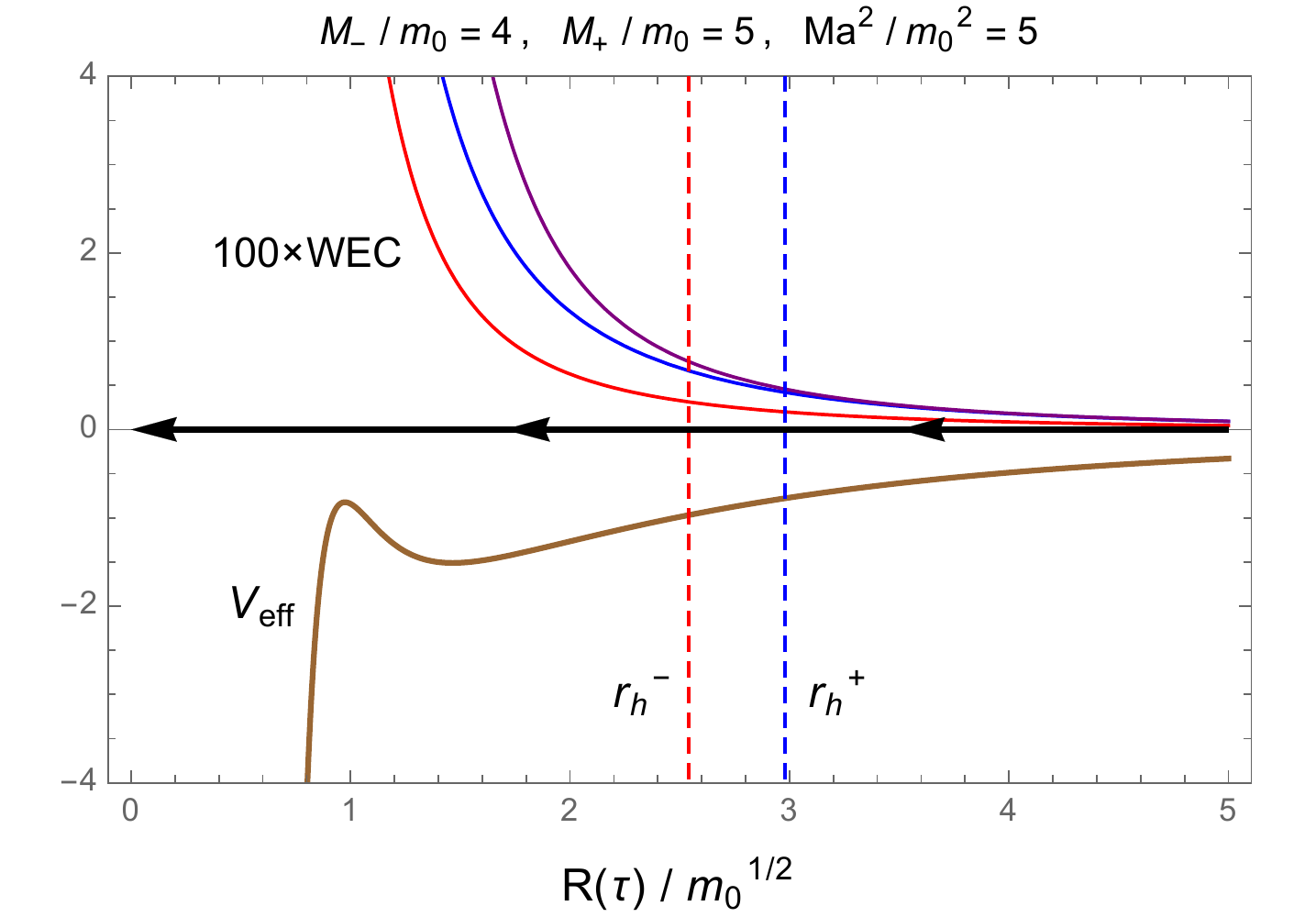}}
\subfigure{\includegraphics[width=0.49\textwidth]{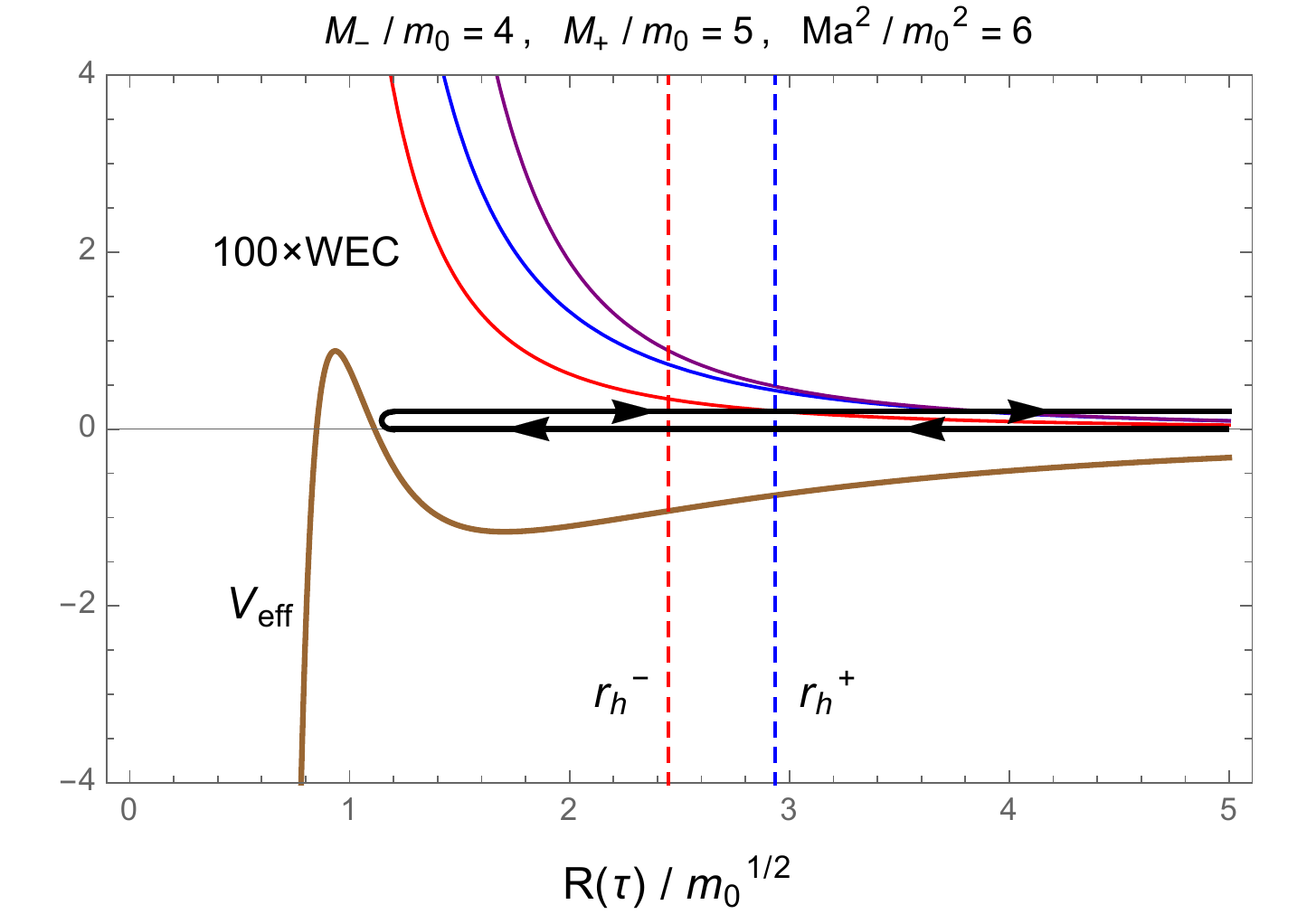}}
\vspace*{2pt}
\caption{Two examples of shell trajectories with zero pressure in five spacetime dimensions (i.e., $N=1$). The left panel describes a plunge and the right panel corresponds to a bounce. The lower (brown) solid curve shows the effective radial potential, Eq.~\eqref{eq:potential}. Classical motion of the shell is allowed whenever $V_{\rm eff}<0$. The curves in the upper half of the plots marked ``100$\times$WEC'' represent the functions defined in Eq.~\eqref{eq:WEC} (rescaled by a hundred for visualization purposes). The fact that they are everywhere positive means that the weak energy conditions are satisfied throughout the whole orbit. The vertical dashed lines mark the location of the interior (red) and exterior (blue) event horizons. For the bounce orbit, the turning point lies inside both horizons. \label{Fig1}}
\end{figure}

Requiring that the shell is initially at rest, inspection of Eq.~\eqref{eq:potential} reveals that this is possible only if $w=0$ and if the shell's rest mass is the difference between the gravitational masses of the exterior and interior geometries, $m_0=M_+-M_-$.
In this case, the effective potential has the following behavior at infinity and near the origin:
\beq
  V_{\rm eff} &\sim& -\frac{M_++M_-}{\cR^{2N}}  \qquad\qquad\qquad\;\;\, {\rm for} \quad \cR \to \infty\,, \\
  V_{\rm eff} &\sim& -\frac{m_0^2}{4\cR^{4N}}\left(1+\frac{2Ma^2}{\cR^{2N+2}}\right) \qquad {\rm for} \quad \cR \to 0\,.  
\eeq
If one turns off the rotation by setting $a=0$ (in addition to $w=0$ and $m_0=M_+-M_-$) it is easy to show that the effective potential is everywhere negative and thus the shell plunges into the black hole already present in the interior spacetime. However, if $a\neq0$ a centrifugal barrier term arises and for sufficiently large rotations the in-falling motion of the shell can be halted and pushed back to infinity, thus leading to a bounce. This situation is signalled by the appearance of a turning point (see Fig.~\ref{Fig1}).

I now comment on the connection between the study of these collapses and the cosmic censorship conjecture. If we were able to find collapses in which 
\begin{enumerate}
\item the interior geometry has an event horizon (and so constitutes a regular initial state from the point of view of an outside observer),
\item the exterior geometry is nakedly singular,
\item the energy conditions are satisfied throughout the evolution,
\end{enumerate}
this situation would configure a stark violation of cosmic censorship.

However, it is easy to show that such a scenario never occurs if the shell has non-negative rest mass, $m_0\geq0$. Indeed, if the interior region has an event horizon, then $g_-(r_-)^{-2}=0$ for some $r_->0$. Then note that
\be
g_+^{-2}(r_-)= g_+^{-2}(r_-)-g_-^{-2}(r_-)= - \frac{2(M_+-M_-)}{r_-^{2N}}= - \frac{2m_0}{r_-^{2N}} \leq 0\,.
\ee
However, since $g_+^{-2}$ is continuous and approaches $1$ as $r\to\infty$ it must have a root at some $r=r_+ \geq r_-$. Therefore, the exterior spacetime also features an event horizon, which is furthermore located outside the event horizon of the interior geometry.

Previous attempts to overspin extremal equal angular momenta black holes with test particles have been shown to comply with cosmic censorship~\cite{BouhmadiLopez:2010vc,Rocha:2014jma}. The above proof, which also applies to the case of a negative cosmological constant, extends those results to the non-perturbative level (and to the non-extremal case) by considering thin shells instead of test particles.

\section{Oscillatory Shells With Rotation\label{sec:oscillatory}}

Ref.~\cite{Delsate:2014iia} constructed stationary thin shell solutions describing shells around rotating black holes in anti-de Sitter (AdS) spacetime. These solutions owe their existence to the combined effect of a confining potential, due to a negative cosmological constant, and a centrifugal barrier due to rotation. For sufficiently large spins (but not too large) one may expect the allowed shell trajectories to be oscillatory. By fine-tuning the free parameters at our disposal one can obtain stationary solutions that satisfy the energy conditions and are, moreover, stable against radial oscillations.

This manuscript is dedicated to the dynamics of rotating thin shells in asymptotically flat spacetimes, which would seem to preclude the above-mentioned possibility. However, a negative pressure component can, to some extent, play the role of the confining potential of AdS. Indeed, if the equation of state parameter is negative, more specifically if $-2N/(2N+1)<w<0$, then the effective potential approaches unity at infinity. If the centrifugal barrier is sufficiently strong one obtains a radial trajectory oscillating between two turning points\footnote{A similar phenomenon arises even for $w=0$ as long as $m_0>M_+-M_-$.}.

\begin{figure}[t]
\centering
\subfigure{\includegraphics[width=0.49\textwidth]{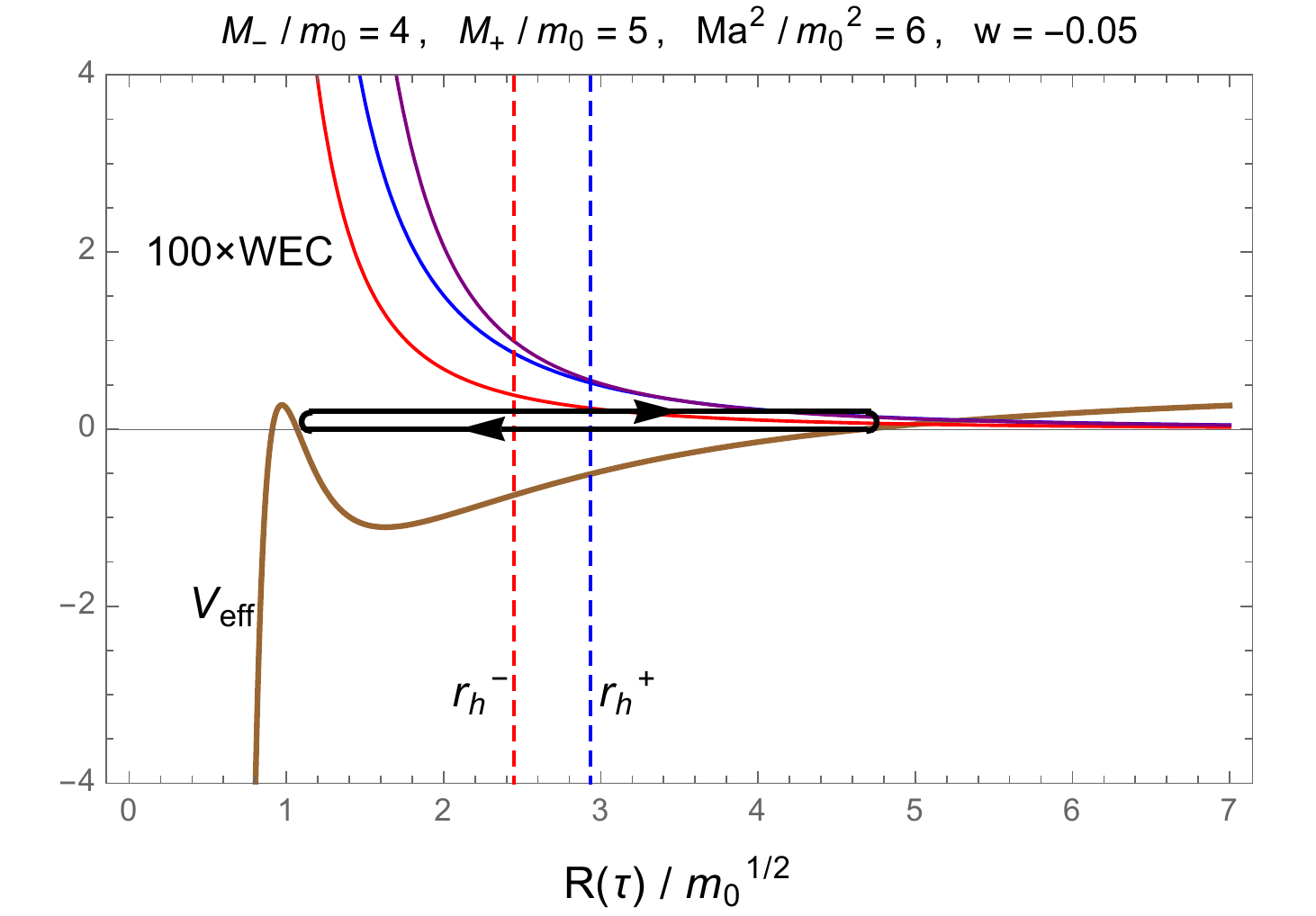}}
\subfigure{\includegraphics[width=0.49\textwidth]{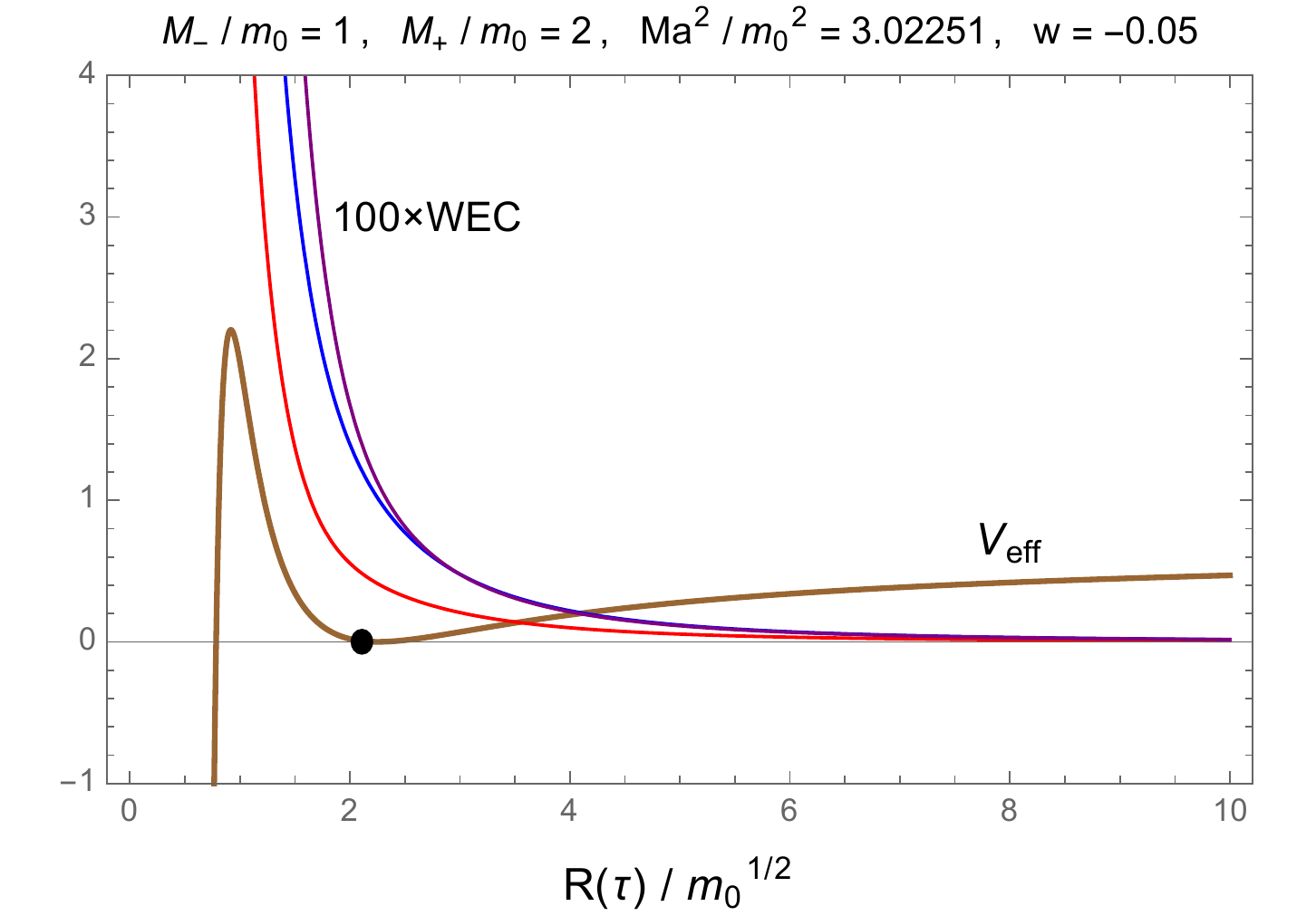}}
\vspace*{2pt}
\caption{Two examples with negative pressure (again in five spacetime dimensions) describing an oscillatory shell (left panel) and a stationary shell (right panel). All the quantities represented are as in Fig.~\ref{Fig1}. For the oscillatory orbit, one of the turning points is outside both event horizons while the other turning point lies inside both horizons. For the stationary orbit, there are no horizons present in the global geometry so the spacetime features a naked singularity. \label{Fig2}}
\end{figure}

In Fig.~\ref{Fig2} an explicit example of such an oscillating (and rotating) shell in five spacetime dimensions is presented. Note that for these oscillatory motions one of the turning points lies inside the event horizons. A similar situation has been noted previously for the case of charged, spherically symmetric shells~\cite{Boulware:1973, Gao:2008jy}. The physical meaning of this becomes clear when one considers the maximal extension of these spacetimes: as the shell oscillates, it falls through the event horizon and then emerges back out into a distinct asymptotically flat region.

One might hope that by fine-tuning the parameters it would be possible to find stationary configurations describing matter shells (with negative pressure) surrounding rotating black holes in asymptotically flat spacetimes. It is indeed possible to obtain stationary shells following the same approach as in Ref.~\cite{Delsate:2014iia} but it appears that they are necessarily nakedly singular.

\section{Concluding Remarks\label{sec:conc}}

The genericalness of the initial data is an important ingredient in any statement of the cosmic censorship conjecture. The framework presented in this note is tailored to extend the study of gravitational collapse to rotating spacetimes, thus going beyond spherical symmetry. The construction relies on the consideration of thin shells and can be regarded as a first step to tackle gravitational collapse of stars with rotation.

From a different perspective, the possibility of having stationary shells hovering above an horizon is certainly interesting: these are exact solutions of the field equations describing stationary shells around rotating black holes and, while being far from resembling an accretion disk, they can serve as useful toy models to study (in)stability issues.

\section*{Acknowledgments}

I thank T\'erence Delsate and Raphael Santarelli for previous and ongoing collaboration on this topic.
J.~V.~R. is supported by {\it Funda\c{c}\~ao para a Ci\^encia e Tecnologia} (FCT)-Portugal through Contract No.~SFRH/BPD/47332/2008.


\end{document}